\documentclass[referee]{raa}           

\usepackage{graphicx,times}
\usepackage{natbib}
\usepackage{amssymb,amsmath}
\usepackage{multirow}
\bibpunct{(}{)}{;}{a}{}{,}

\usepackage[pagebackref=false]{hyperref}

\begin{document}

 \title{Advancing Gamma-Ray Burst Identification through Transfer Learning with Convolutional Neural Networks}

 \volnopage{ {\bf 20XX} Vol.\ {\bf X} No. {\bf XX}, 000--000}
   \setcounter{page}{1}

\author{
   Peng Zhang\inst{1,2}, 
   Bing Li\inst{2,3}, 
   Ren-zhou Gui\inst{1}, 
   Shao-lin Xiong\inst{2,4}, 
   Yu Wang \inst{5,6,7},
   Yan-qiu Zhang\inst{2,8}, 
   Chen-wei Wang\inst{2,8}, 
   Jia-cong Liu\inst{2,8}, 
   Wang-chen Xue\inst{2,8},
   Chao Zheng\inst{2,8},
   Zheng-hang Yu\inst{2,8},
   Wen-long Zhang\inst{2,9}
}

\institute{ 
College of Electronic and Information Engineering, Tongji University, Shanghai 201804, China; rzgui@tongji.edu.cn\\
\and
Particle Astrophysics Division, Institute of High Energy Physics, Chinese Academy of Sciences, Beijing 100049, People's Republic of China; libing@ihep.ac.cn\\
\and
Guangxi Key Laboratory for Relativistic Astrophysics, Nanning 530004, People's Republic of China\\
\and
Key Laboratory of Particle Astrophysics, Chinese Academy of Sciences, Beijing 100049, China; xiongsl@ihep.ac.cn\\
\and
ICRA, Dip. di Fisica, Sapienza Universit\`a  di Roma, Piazzale Aldo Moro 5, I-00185 Roma, Italy
\and
ICRANet, Piazza della Repubblica 10, 65122 Pescara, Italy
\and
INAF -- Osservatorio Astronomico d'Abruzzo, Via M. Maggini snc, I-64100, Teramo, Italy
\and
University of Chinese Academy of Sciences, Beijing 100049, China
\and
School of Physics and Physical Engineering, Qufu Normal University, Qufu, Shandong 273165, China\\
}

\abstract{
The Rapid and accurate identification of Gamma-Ray Bursts (GRBs) is crucial 
for unraveling their origins. However, current burst search algorithms 
frequently miss low-threshold signals or lack universality for observations. 
In this study, we propose a novel approach utilizing transfer learning 
experiment based on convolutional neural network (CNN) to establish a 
universal GRB identification method, which validated successfully using
GECAM-B data. By employing data augmentation techniques, we enhance 
the diversity and quantity of the GRB sample. We develop a 1D CNN model 
with a multi-scale feature cross fusion module (MSCFM) to extract 
features from samples and perform classification. The comparative
results demonstrated significant performance improvements following 
pre-training and transferring on a large-scale dataset. 
Our optimal model achieved an impressive accuracy of 96.41\% on the 
source dataset of GECAM-B, and identified three previously undiscovered 
GRBs by contrast with manual analysis of GECAM-B observations. 
These innovative transfer learning and data augmentation methods presented 
in this work hold promise for applications in multi-satellite exploration 
scenarios characterized by limited data sets and a scarcity of labeled 
samples in high-energy astronomy.
}

   \authorrunning{Zhang et al. }       
   \titlerunning{Advancing GRB Identification through Transfer Learning }  
   \maketitle

\section{Introduction}           
\label{sect:intro}
Gamma-ray Bursts (GRBs) release immense amounts of energy in a remarkably
short time frame, predominantly in the form of gamma rays band. 
To date, the origins of gamma-ray bursts remain unknown and the
classification of GRBs still is a crucial research endeavor, making them 
a continuing research hotspot in the field of astronomy 
\citep{Zhang2011CRP,KumarZhang2015PhR,Meszaros2019MmSAI,
Sun2023Arxiv,Wang2024Arxiv}.
As more high-quality space probes for exploring GRBs are in orbit, e.g., Swift/BAT \citep{swift_bat}, Fermi/GBM \citep{fermi_gbm}, GECAM/GRD
~\citep{gecam}, SVOM/GRM \citep{svom}, rapid and accurate identification of
gamma-ray burst events becomes especially important.
And with the leapfrogging of multi-messenger, multi-band astronomy, the rapid
identifying GRBs from detectors of space-based missions is advantageous for
guiding other telescopes to conduct joint or follow-up observations for 
studying afterglows, locating host galaxies, hunting other counterparts and
intrinsic information. The light curves of GRBs exhibit irregular and 
multi-peaked, and these shapes undoubtedly contain wealth of physical 
characteristics. 
Traditional burst search algorithms capture signals in multiple time-bin 
light curves with signal-to-noise ratios (SNRs) above a predefined threshold 
exceeded the background \citep{fermi_blind_search,fermi_target_search,search_grb_hxmt_cai}.
However, these methods are challenging to accurately estimate the background
level and set appropriate thresholds.

As a sub-field of artificial intelligence, Machine Learning (ML) is routinely 
used today for a wide range of purposes. ML techniques are exploit only to 
process, visualize, and make predictions from big data, but also to make 
data-driven discoveries. The operational capability of ML algorithms for 
autonomously learning, adapting, and improving from data were impressively 
demonstrated. The information content of a dataset is organised, and the 
knowledge of internal samples could be extracted through unsupervised, 
semi-supervised, or supervised learning approaches. 
The applications of such methods are becoming commonplace and growing rapidly 
in modern domain of astronomy and astrophysics. 
For a graphical overview of heavy use in extragalactic astronomy see 
\citep{Fotopoulou2024AC}.
ML along with its subset, deep learning (DL), have revolutionized the astronomy 
domain by providing powerful tools for data analysis and pattern recognition 
in recent years, and the application of such techniques in astronomical 
researches have gained significant attention and 
traction \citep{BaronDalya2019arX, Djorgovski2022arX}.
The utilization of machine learning methods showed great potential in enhancing
the detection and classification of fast transient events \citep{duBuisson2015MN,
Wagstaff2016PASP, Sooknunan2021MN, Alves2022ApJS}.

ML has been also applied as a method for the discovery of GRBs experimentally. 
By combining density-based spatial clustering and dynamic time warping for 
template matching, \cite{relate_search_GRB_DTW} developed a ML algorithm for
automated detection of GRB-like events using the AstroSat/CZTI data, 
which demonstrated various resourcefulness of this method for robust GRB 
identification. As a specialized type of DL algorithm, Convolutional Neural 
Networks (CNNs) mainly designed for tasks that necessitate object recognition,
including image classification, object detection, and segmentation (for review 
see \citep{Alzubaidi2021ReviewOD, Taye2023TheoreticalUO}. 
CNNs are employed in a variety of practical scenarios in astronomy for working 
and processing  with variety data from observations, such as light curves, 
images, and spectra. They have brought about revolutionary changes in data 
analysis by efficiently automating the extraction of complex features from 
astronomical datasets.
\cite{DL_detect_GRB_intensity_map} develop an automated real-time analysis
pipeline that consisted of a new approach of CNN to classify the intensity 
maps acquired by AGILE-GRID. In a recent paper, a quantum CNN was implemented 
to detect GRBs from sky maps or light curves of this telescope, 
which achieved considerable improvements of accuracy of recognition 
\citep{Rizzo2024arX}. These improving of GRB detection capability and 
effective were demonstrated obviously. 
\citet{DL_autodencoder_detect_grb} apply a CNN-based auto-encoder to 
reconstruct background light curves and detect GRBs by exceeding the threshold 
reconstruction error. Their approach successfully identified 72 GRBs that 
were not listed in the AGILE catalog. In addition, A deep neural network 
was trained with a vast amount of AGILE orbital and attitude parameters to 
predict the background count rates, yielding promising results and enabling 
the development of an anomaly detection method for detecting GRBs \citep{ParmiggianiarX240402107P}. 
\citet{nn_search_long_faint_burst} employ neural networks to estimate the 
long-term background level using orbital and spatial environmental information.
Combining the trigger search algorithm based on SNR, seven suspected long 
faint bursts were identified. 
The training of DL models rely on a significant amount of well-annotated data, 
while the detected bursting events of astronomical transients are exceedingly 
limited. A wealth of experimental or validation operations of such methods 
are indeed available at present. However, it is undeniable that the dataset 
embedded with rich features and diversity is still limited, which still leads 
to many questionable applications and results.

Transfer learning (TL) is proposed to alleviate the issue of insufficient 
generalization capability of models due to limited samples. 
TL leverages knowledge learned from source task to improve learning and 
generalization on target task, offering benefits such as faster convergence, 
improved performance, and reduced data requirements \citep{Pan2010IEEE,Yosinski2014Arxiv,Kim2021Arxiv}. 
DL algorithms for galaxy morphological classification have achieved 
remarkable success, but their reliance on large labelled training samples 
raises the problem of transfer ability to new datasets \citep{Barchi2020AC}.
\cite{Dominguez2019MNRAS} demonstrates that transfer learning, by pre-training 
CNN model with Sloan Digital Sky Survey data and adapting them to Dark Energy 
Survey data through parameter fine-tuning. 
This method significantly improves classification accuracy and completeness, 
reducing the required training sample size.
\cite{Awang2020Galaxies} assesses the efficacy of CNN for planetary nebulae 
classification, focusing on distinguishing such objects from their morphology. 
Utilizing transfer learning with pre-trained algorithms, 
the study achieved high success in differentiating True PNe without parameter fine-tuning. 
\cite{Cavuoti2024AA} presents a novel outlier detection method in astronomical
time series, highlighting the effectiveness of transfer learning using a CNN 
model pre-trained on ImageNet dataset. 
Application of transfer learning to VLT Survey Telescope data demonstrates 
its practicality in artifact identification and removal, showcasing its 
potential for enhancing data quality and reliability. 
The transfer learning alleviates the challenges posed by limited labeled 
data and model generalization. 
It also demonstrates its capability to enhance the reliability, stability, 
and accuracy of models across various astronomical tasks. 
Current deep learning methods for GRB searches are customized to a single 
satellite, and the effective transfer to satellites with limited observation 
is equally crucial.

In our previous studies, \citet{DL_identify_grb_by_peng} adopt multilayer 
CNNs to distinguish GRBs from observation data of Fermi/GBM. As input data, 
the count maps contain abundant features, including light curves and spectral 
information in a wide-bandwidth, which were successfully utilized to train 
models. These processing methods achieved high accuracy and demonstrated 
their effectiveness through feature visualization and classification.
Applying the optimal model to one year observations, it achieves results 
comparable to manual analysis,and furthermore, it shows remarkable capacity 
for identification of sub-threshold GRBs. 
These outstanding results are mainly attributed to the relatively large 
labeled samples of GRBs observed by Fermi/GBM.
Undoubtedly, the current limited quantity of labeled samples of GRBs makes 
it challenging to fully harness the potential of neural networks. 
Especially for telescopes with a short time in orbit, e.g., ravitational wave
high-energy Electromagnetic Counterpart All-sky Monitor (GECAM, \cite{gecam}), The Space-based multi-band astronomical Variable Objects Monitor 
(SVOM, \cite{Atteia2022IJMPD} ) 
Hence the transfer learning caught our attention due to its ability to address 
data limitations by leveraging knowledge learned from related tasks or domains.
The main datasets employed in our study comes from the gamma-ray detectors 
 (GRDs) onboard GECAM-B satellite that in orbit for three years.
As a purpose for exploring how to transfer models trained on large-scale datasets, we utilized source datasets referenced to \citet{DL_identify_grb_by_peng}.
The dataset, data augmentation, and data pre-processing are presented 
in Section \ref{sec:dataset}. 
The Section \ref{sec:methods} shows transfer learning model structure, 
pre-training and fine-tuning in detail.
The performance of the model and the recovered GRBs are shown in Section \ref{sec:result}.
Section \ref{sec:discussion} discusses the performance of the model, the practical implications of the new approach.

\section{Dataset}
\label{sec:dataset}
Benefiting from the wide energy band, large field of view, and high sensitivity 
of gamma-ray detectors of GRD, the GECAM-B satellite has detected more than 
300 GRBs since its launch in December 10, 2020. 
These 25 GRDs have demonstrated strong capabilities in detecting gamma rays and 
particles, as well as determined the locations of GRBs roughly \citep{gecam_grd}. 
We constructed a target dataset based on the observed data from GECAM-B, 
which includes categories for both GRB and non-GRB samples.
In order to obtain valid GRB category samples, we screened the GRBs detected 
by the GECAM-B/GRD. Through manual analysis, there are 219 GRBs 
with precise location between 01/01/2021 and 02/01/2024. 
Corresponding to these GRBs that triggered by no fewer than three detectors, 
along with their location and direction of every GRD detector, we calculated 
the incidence angles of every GRB source for detectors.
The extraction of light curves from GECAM data is facilitated through GECAMTools\footnote{\url{https://github.com/zhangpeng-sci/GECAMTools-Public/}}.
Considering the requisite for high statistical light curves of GRB samples,
we only extracted event data from the three detectors that with relatively 
small incidence angles for each event. This process resulted in 657 light 
curves that were collected to augment for GRB category samples. 
Each GRB sample was set to a duration of 120 seconds, with a minimum of 10 
seconds of background data before and after the burst to ensure the integrity 
of the burst shape.
For the non-GRB category data from GCEAM-B, we randomly sampled 1,500 daily 
observational files from non-burst time periods. 
From each file, twenty 120-second segments were extracted without overlap 
for additional analysis, this gives us 11,000 non-GRB samples. 
The non-GRB category data are shown in the table \ref{table:dataset}.
We select photons recommended in all gain types without distinguishing 
in the high gain or low gain. 
These photons are utilized to generate light curves with a time bin of 
64\,ms from both GRB and non-GRB data.

To facilitate the pre-training of our model, we build the Fermi/GBM dataset
(source dataset). As of July 2023, Fermi/GBM has manually identified over 
3500 GRBs. Following the GRB category and non-GRB category data extraction 
methods presented in \citep{DL_identify_grb_by_peng}, 
we have made updates to the time range of the data selection from July 14, 
2008 to June 31, 2023. We initially achieved 6,189 GRB samples and 108,000
non-GRB samples. 
Relatively speaking, a small time-bin light curve provides deeper and detailed
information that is more favorable identifying GRBs \citep{DL_identify_grb_by_peng}. 
The energy band of all light curves are binned into 9 universal bins
(25-50 keV, 50-100 keV, 50-300 keV, 100-300 keV, 100-500 keV, 100-900 keV, 
300-500 keV, 300-900 keV, and 500-900 keV), refereed to the trigger search 
algorithm of the Fermi/GBM \citep{fermi_gbm_fouth_catalog,search_grb_hxmt_cai}. 
The majority of GRB photons are distributed across these energy bands, 
and the energy band binning enhances the intensity of the burst signal in 
the light curves.

Normalization and standardization are common techniques used in data 
pre-processing to scale and transform the features of a dataset. 
Normalization involves rescaling the values of a feature to a specific range, 
typically between 0 and 1. 
This is achieved by subtracting the minimum value of the feature and dividing 
by the range (maximum value minus minimum value). 
On the other hand, standardization transforms the values of a feature to have 
a mean of 0 and a standard deviation of 1. 
This process involves subtracting the mean value of the feature from each 
value and dividing by the standard deviation. 
Both normalization and standardization are employed to ensure that the 
features in a dataset are on a similar scale. 
The ResNet model, as proposed by \cite{model_resnet}, is a widely used 
model in deep learning. 
In this study, we employed the ResNet model to evaluate and compare 
the impact of various pre-processing methods on performance. 
The results of the performance evaluation on the source test set 
using these two methods are summarized in Table \ref{table:data_preprocess}. 
Given that standardizing samples leads to the highest accuracy for each 
energy band, 
we have decided to adopt standardizing method for pre-processing the samples 
in source and target dataset.

Data augmentation has emerged as a widely adopted technique for improving 
the generalization capabilities of deep learning models.
The diversity of the data influences the model's ability to generalize.
Additionally, overfitting or underfitting occurs when a model has learned 
limited flexibility from the dataset, either due to a lack of features or 
excessive regularization.
One effective way to prevent these issues is by considering not only the 
quantity aspect but also the similarity and diversity aspects of data.
While it may not be a guarantee, training with a larger quantity of data can 
assist deep learning algorithms in better detecting signals.
In this study, we experimentally present a data augmentation technique for 
the light curves of GRB category data. The GRB signals utilized in this 
dataset have been manually analyzed, resulting in overwhelmingly high 
SNR throughout the distribution. 
This prevalence of high SNR signals has led to a scarcity of samples with 
relatively lower SNRs, and a notable shortage of of sub-threshold 
samples within our GRB category data.

\begin{figure}
\centering
\includegraphics[width=0.9\textwidth, angle=0]{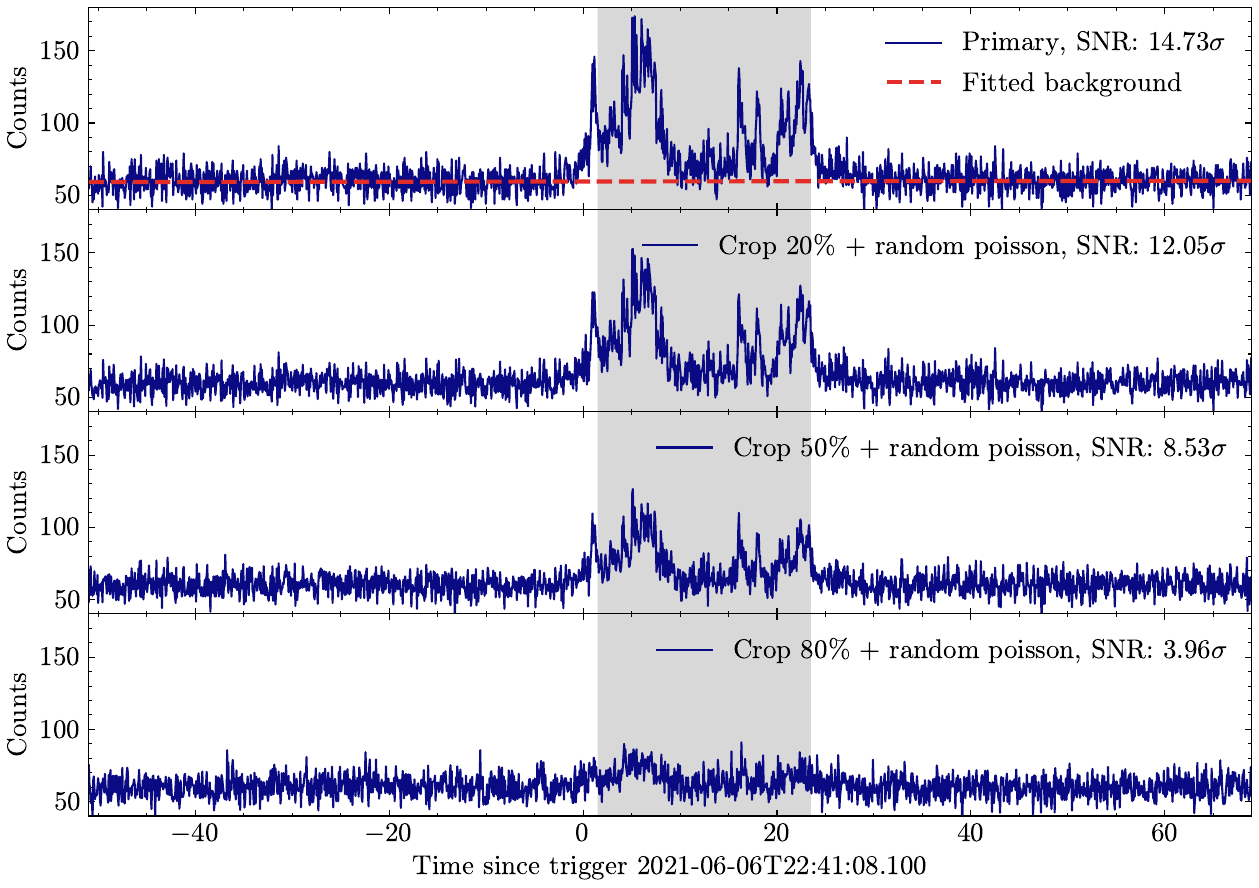}
\caption{Illustration of our data augmentation. 
The top panel displays the the primary light curve of GRB 210606B in total 
energy channel (detected by GECAM-B/GRD02). The SNR for this sample is 
14.73 $\sigma$. The red dotted line represents the background fitted with a 
polynomial function, and the background level approximately equal 
to 52 counts per bin. The shaded region corresponds to the primary GRB region.
The bottom three panels respectively show the light curves after reducing 
the counts of each bin by proportionally about 20\%, 50\% and 80\%, followed
by back-filling the background by overlaying counts sampled from a Poisson 
distribution with a mean value equal to the reduced background counts.
Their respective SNRs are marked.}
\label{fig:crop_example}
\end{figure}

Our objective is to create a substantial amount of mocked GRB signals with 
lower SNRs and sub-threshold levels based on the existing light curves of GRBs.
We consider an idea in which the SNR of each individually primary GRB is 
randomly decreased multiple times by reducing counts in each bin according 
to a stochastic proportion.
This idea bears a resemblance to the scaling or resizing method of affine transformations used for augmenting image data. Despite the simplicity of 
these affine techniques, several studies have indicated that they are 
highly effective in various machine learning tasks \citep{Mumuni2022DataAA, KumarT2023arXiv230102830K, WangZ2024arXiv240509591W}.
The negative samples (i.e. non-GRBs) we utilize that are extracted directly 
from observations in non-burst regions, representing the background level of 
the detectors in orbit. Therefore, the method of randomly reducing counts on 
a random scale may pose a challenge, as it could lower the background level 
to an unrealistic extent, potentially complicating signal detection and 
identification.
Our data augmentation approach involves reducing the counts of each individual 
GRB light curve while maintaining the background level, as illustrated in 
Figure \ref{fig:crop_example}.

To implement this method, we initially select the stable background level 
both in the preceding and the subsequent of the burst and derive the 
background light curve $Lc_{bg}$ by fitting the background using a 
polynomial function. 
Subsequently, we subtract counts from the primary light curve $Lc$ based on 
a specified $crop\_factor$, and then reintroduce the corresponding 
Poisson-sampled background ratio.
\begin{equation}
Lc_{new}=Lc \times (1-crop\_factor) + random\_possion( Lc_{bg} \times crop\_factor )
\end{equation}
where $Lc_{new} $ represents the newly generated light curve.  
The $crop\_factor$ for each data augmentation is randomly sampled from a 
log-normal distribution with a mean of -2 and a variance of 0.6.

To enhance the diversity of the training set, we augmented the GRB category 
data either GRBs selected form GECAM-B/GRD or GRBs selected form Fermi/GBM.
The former supplies initial 657 light curves of GRBs, and the latter provides
initial 6189 light curves of GRBs.
After data augmentation handling discussed above, they were successfully multipled about 23 times, resulting in a sizeable GRB category samples both on target dataset and source dataset finally.
The GRB category samples and non-GRB category samples are divided into respective training set, validation set, and test set based on time period. This is very helpful in avoiding data confusion. The ratio of positive and negative sample numbers in these three data sets is roughly comparable and reasonable. Details are described in the table \ref{table:dataset}.

\begin{table}
\bc
\begin{minipage}[]{100mm}
\caption[]{Description of the target and source dataset.\label{table:dataset}}\end{minipage}
\setlength{\tabcolsep}{1pt}
\small
 \begin{tabular}{ccccc}
  \hline\noalign{\smallskip}
Dataset & Partition & Nu. of GRB & Nu. of non-GRB & Data Period Definition (UTC)\\
  \hline\noalign{\smallskip}
\multirow{3}{*}{Target dataset} & Training set & 10005 & 10000 &  01/01/2021 - 11/30/2022 \\ 
& Validation set & 108 & 500 & 01/12/2023 - 05/31/2023 \\
& Test set & 114 & 500 & 06/01/2023 - 31/01/2024 \\
   \noalign{\smallskip}\hline
\multirow{3}{*}{Source dataset} & Training set & 105213 & 100000 &  07/14/2008 - 31/12/2016 \\ 
& Validation set & 2333 & 4000 & 01/01/2017 - 12/31/2019 \\
& Test set & 2776 & 4000 & 01/01/2020 - 06/31/2023 \\
  \noalign{\smallskip}\hline
\end{tabular}
\ec
\end{table}

\begin{table}
\bc
\begin{minipage}[]{100mm}
\caption[]{The difference of the data pre-process methods.\label{table:data_preprocess}}\end{minipage}
\setlength{\tabcolsep}{1pt}
\small
 \begin{tabular}{cccccc}
  \hline\noalign{\smallskip}
Model & Pre-process & \textit{Accuracy} (\%) & \textit{Precision} (\%) & \textit{Recall} (\%) & \textit{F1-score} (\%)\\
  \hline\noalign{\smallskip}
\multirow{4}{*}{ResNet} & Norm & 94.86   & 98.61   & 90.23   & 94.24 \\ 
& Norm each channel   & 95.82   & 98.29   & 91.39   & 94.71 \\ 
& Standard   & 96.75   & 98.22   & 93.76   & 95.94 \\ 
& Standard each channel & \textbf{97.15}   & \textbf{98.42}   & \textbf{94.56}   & \textbf{96.45} \\
  \noalign{\smallskip}\hline
\end{tabular}
\ec
\tablecomments{0.86\textwidth}{Bold text represents that the model performs optimally on that metric.}
\end{table}

\section{Model architecture, training, and extension}
\label{sec:methods}

\subsection{Architecture of Neural Networks}
\label{sec:nework_architectures} 
\begin{figure}[h]
	\begin{minipage}{\linewidth}
		\vspace{0pt}
		\centerline{\includegraphics[width=\textwidth]{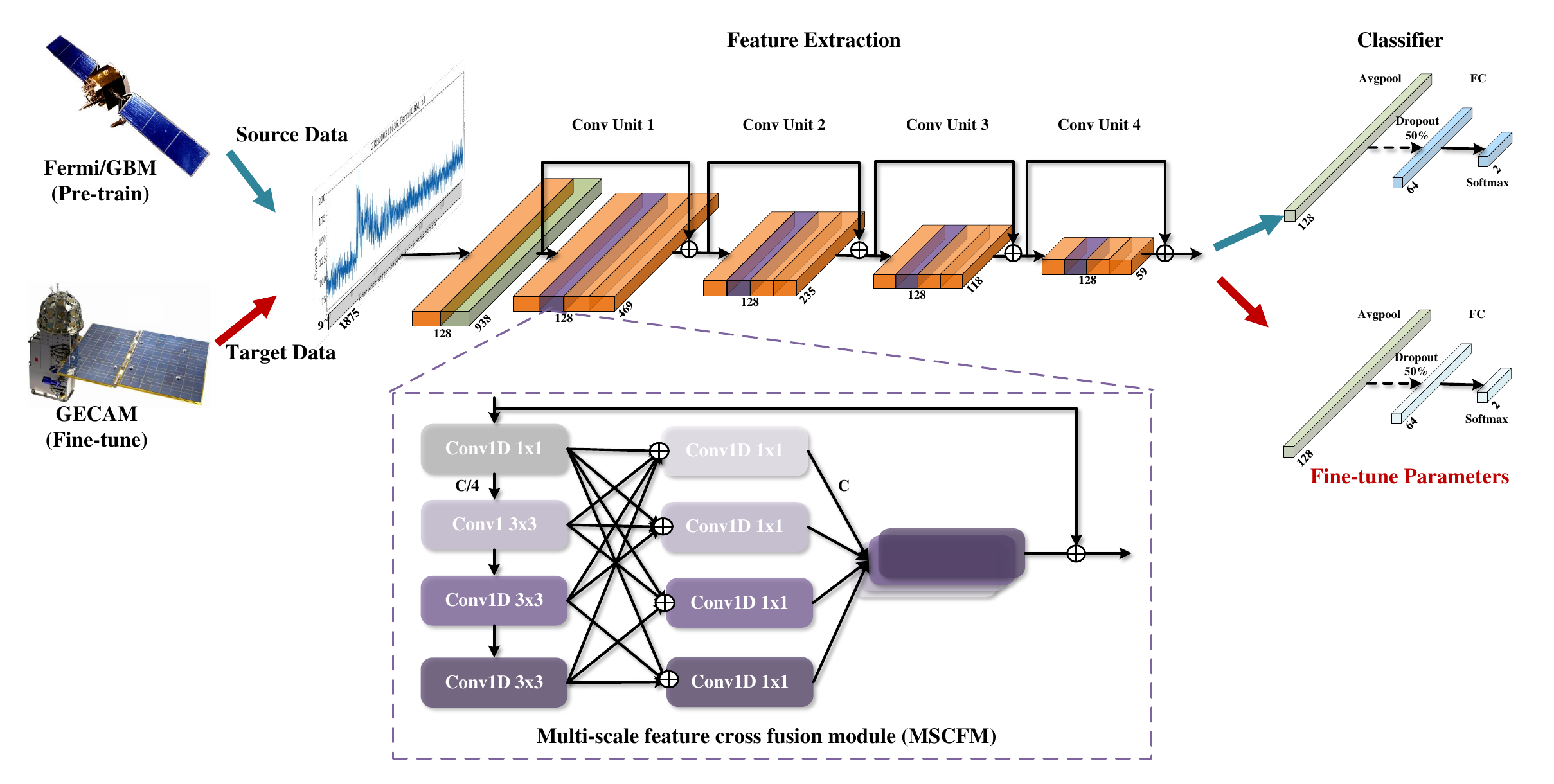}}
	\end{minipage}
	\caption{Schematic diagram of model architecture and process of layers' transfer. The numbers in the convolutional units indicate changes of the size and dimensions of the features. The purple module represents the MSCFM module, with the detailed implementation of this module shown in the sub-graph at the bottom.}
	\label{fig:network_architectures}
\end{figure}

Each sample within our dataset comprises light curves across nine distinct energy bands, 
making them suitable for treatment as time series data using a 1D convolutional neural network. 
Our model's foundation is based on ResNet \citep{model_resnet}, which leverages 
a series of four convolutional units (Conv Units) to extract features from the data. 
ResNet introduces residual connection to address the issue of gradient vanishing in deep neural networks.
Following feature extraction by the convolutional layers, the extracted features are passed through two fully connected layers (FC) for classification after global average pooling. 
To prevent over-fitting, a dropout layer with a 50\% probability is incorporated 
between the global average pooling and the fully connected layer. 
The comprehensive architecture of our model is depicted in Figure \ref{fig:network_architectures}.

Building upon the ResNet architecture, we introduce a novel 
multi-scale feature cross fusion module (MSCFM) designed to 
enhance the representation of multi-scale features. 
Motivated by the distinct behavior of GRB light curves across various time scales, 
we incorporate MSFM modules to extract and fuse features at different scales 
within our deep learning model. 
This approach enables the model to capture the detailed characteristics 
of the burst event across multiple scales, enhancing robustness and accuracy. 
To facilitate multi-scale feature extraction, we initially employ 
a $1\times1$ convolutional layer to reduce feature dimensionality, 
thereby reducing the number of parameters and computational load. 
Subsequently, we stack three convolutional layers with a kernel size 
of $1\times3$ to extract features at larger scales, 
leveraging the increased receptive field of deeper convolutions. 
The features extracted at three distinct scales are then cross-fused with 
the primary features, and the resulting cross-fused features are restored 
to their original dimensions using a $1\times1$ convolutional layer. 
By incorporating these cross-fusion features into the original features 
through a residual structure, we obtain multi-scale cross-fusion features. 
The MSCFM module is integrated into each convolutional unit of the model, 
as indicated by the purple module in Figure \ref{fig:network_architectures}. 
This innovative approach enhances the model's ability to capture and leverage 
multi-scale information, ultimately improving its performance 
in GRB identification tasks.

Optimizing hyper-parameters in deep learning models is essential for enhancing 
model performance and generalization by improving data fitting and preventing over-fitting. 
By adjusting hyper-parameters, we could expedite model training and ensure optimal performance.
In our work, we conduct a detailed analysis to evaluate the impact of different hyper-parameter 
choices on the test set of the source dataset, as referenced in \citep{dl_technosignatures}. 
Through this analysis, we identify the hyper-parameter set that yields the best-performing models. 
The hyper-parameter configurations that result in the best models are highlighted in bold in Table \ref{table:hyper_parameter_select}. 
This systematic evaluation of hyper-parameters allows us to improve our model and enhance its effectiveness in identifying GRBs.

\begin{table}
\bc
\begin{minipage}[]{100mm}
\caption[]{Hyper-parameter selection.\label{table:hyper_parameter_select}}\end{minipage}
\setlength{\tabcolsep}{1pt}
\small
 \begin{tabular}{cc}
  \hline\noalign{\smallskip}
Parameters & Values\\
  \hline\noalign{\smallskip}
Number of ConvUnit & 2, \textbf{4}, 8, 16 \\
Number of Conv in ConvUnit & 1, 2, \textbf{3}, 4 \\
First ConvUnit filter size & 64, \textbf{128}, 256, 512 \\
Second ConvUnit filter size & 64, \textbf{128}, 256, 512 \\
Third ConvUnit filter size & 64, \textbf{128}, 256, 512 \\
Fourth ConvUnit filter size & 64, \textbf{128}, 256, 512 \\
Norm function & InstaceNorm, \textbf{BatchNorm} \\
Activation function & Sigmoid, \textbf{Relu} \\
FC neurons & 32, \textbf{64}, 128 \\
Dropout rate & 0.3, \textbf{0.5}, 0.8 \\
  \hline\noalign{\smallskip}
Initial learning rate  & 1e-3, 1e-4, \textbf{1e-5}, 1e-6 \\
Batch size & 512, \textbf{1024}, 2048\\ 
Patience of reduce learning rate & 5, \textbf{10}, 15\\
Patience of early stop & 10, \textbf{20}, 40, 60\\
  \noalign{\smallskip}\hline
\end{tabular}
\ec
\tablecomments{0.86\textwidth}{Bold text represents the selected hyper-parameter.}
\end{table}

\subsection{Model Pre-training on Fermi/GBM dataset}
\label{sec:train_model}
To initiate the training of our deep learning model, we initialize the model parameters 
using the truncated normal distribution method proposed by \citep{param_he_normal}. 
The batch size for model training is set to 1024, and the cross-entropy loss function 
is employed as the supervised model's loss function to measure the disparity between 
the predicted and actual label of the data. 
For model optimization, we utilize the Adam optimizer \citep{adam_optimiser}, 
which updates the model parameters to minimize the loss during neural network training. 
The initial learning rate of the model is established at 1e-5.
In the training process, if the validation loss fails to decrease for 10 consecutive epochs, 
we reduce the learning rate by a factor of 2. 
To prevent over-fitting, we implement an early stopping mechanism, 
terminating the training process when the validation accuracy does not improve for 20 consecutive epochs. 
The selection of hyper-parameters such as batch size, initial learning rate, and early stopping criteria 
are detailed in Table \ref{table:hyper_parameter_select}. 
The parameters from the epochs with the highest accuracy in the validation set are 
utilized as the optimal model parameters.
Our model is implemented using the Pytorch framework on a single GPU (NVIDIA RTX-4090). 
The evaluation metrics for our model, including 
\textit{Accuracy}, \textit{Precision}, \textit{Recall}, and \textit{F1-score}, 
are consistent with those described in \citep{DL_identify_grb_by_peng}.

\begin{table}
\bc
\begin{minipage}[]{100mm}
\caption[]{Model performance on Fermi/GBM dataset.\label{table:model_perform_gbm}}\end{minipage}
\setlength{\tabcolsep}{1pt}
\small
 \begin{tabular}{ccccc}
  \hline\noalign{\smallskip}
Model & \textit{Accuracy} (\%) & \textit{Precision} (\%) & \textit{Recall} (\%) & \textit{F1-score} (\%)\\
  \hline\noalign{\smallskip}
ResNet & 97.15   & \textbf{98.42}   & 94.56   & 96.45 \\
ResNet+MSCFM & \textbf{97.37}   & 97.96   & \textbf{95.56}   & \textbf{96.75} \\
  \noalign{\smallskip}\hline
\end{tabular}
\ec
\tablecomments{0.86\textwidth}{Bold text represents that the model performs optimally on that metric.}
\end{table}

\subsection{Transfer Model to GECAM dataset}
\label{sec:transfer_model}
To enhance the performance of our model in identifying GRBs, 
we initially binned the energy bands of the light curves into nine distinct bands for pre-training on the GBM dataset. 
This approach enables us to easily transfer the model to the GECAM satellite 
by binning the energy ranges of the observed light curves to the similar energy range.
Subsequently, we collected GECAM-B observations to construct the target dataset. 
Leveraging the pre-trained model on the source dataset, we fine-tuned the model using the target dataset. 
During fine-tuning, we opted to freeze all parameters of the convolutional layer 
and solely adjusted the parameters of the fully connected layer. 
This indicates that we maintain the feature extraction capability of the model 
and only update the classifier to suit the new data.
The learning rate was set to a small value of 1e-6 to facilitate the fine-tuning process, 
ensuring gradual adjustments to the model's performance.
The training settings for the model remained consistent with Section \ref{sec:train_model}. 
Furthermore, we conducted an investigation to ascertain whether the model effectively 
extracted the key features of the GRBs. 
To elucidate the model's decision-making process, we employed the 
Grad-CAM (Gradient-weighted Class Activation Mapping) method, as detailed in \citep{model_grad_cam}. 
This method visually highlights the features that the model focuses on during prediction, 
providing valuable insights into the model's decision-making rationale.

\begin{table}
\bc
\begin{minipage}[]{100mm}
\caption[]{Model performance on GECAM dataset.\label{table:model_perform_gecam}}\end{minipage}
\setlength{\tabcolsep}{1pt}
\small
 \begin{tabular}{ccccccc}
  \hline\noalign{\smallskip}
Model & Pre-train & Fine-tune & \textit{Accuracy} (\%) & \textit{Precision} (\%) & \textit{Recall} (\%) & \textit{F1-score} (\%)\\
 & dataset & dataset &  &  & & \\
  \hline\noalign{\smallskip}
  \multirow{3}{*}{ResNet-MSCFM} & GECAM & - & 95.44   & \textbf{95.74}   & 78.94   & 86.54 \\
& Fermi/GBM & - & 96.25   & 90.99   & 88.59   & 89.77 \\
& Fermi/GBM & GECAM & \textbf{96.41}   & 91.07   & \textbf{89.47}   & \textbf{90.26} \\
  \noalign{\smallskip}\hline
\end{tabular}
\ec
\tablecomments{0.86\textwidth}{Bold text represents that the model performs optimally on that metric.}
\end{table}

\begin{figure}[htbp]
	\centering
	\begin{minipage}{0.49\linewidth}
		\centering
		\includegraphics[width=\linewidth]{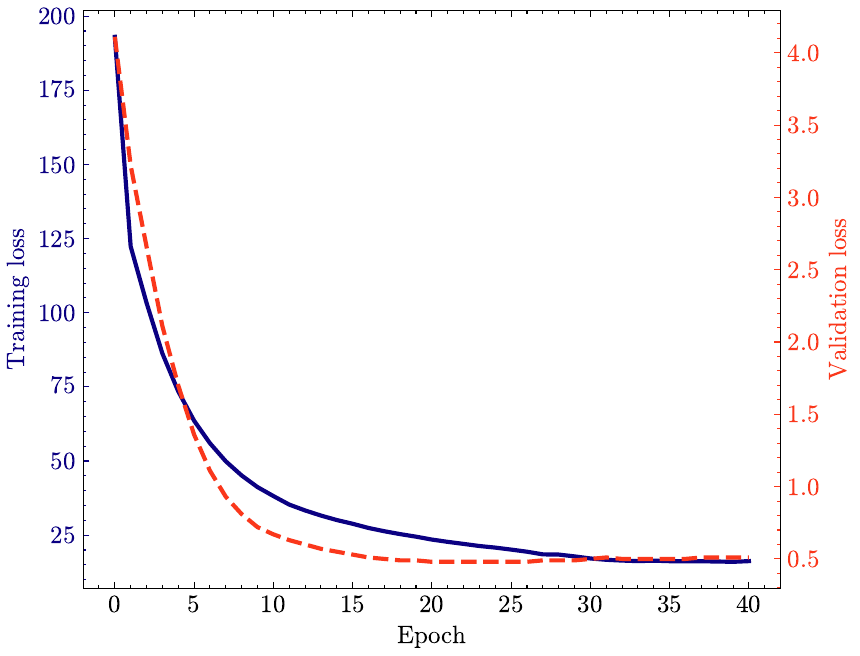}
	\end{minipage}
	\begin{minipage}{0.49\linewidth}
		\centering
		\includegraphics[width=\linewidth]{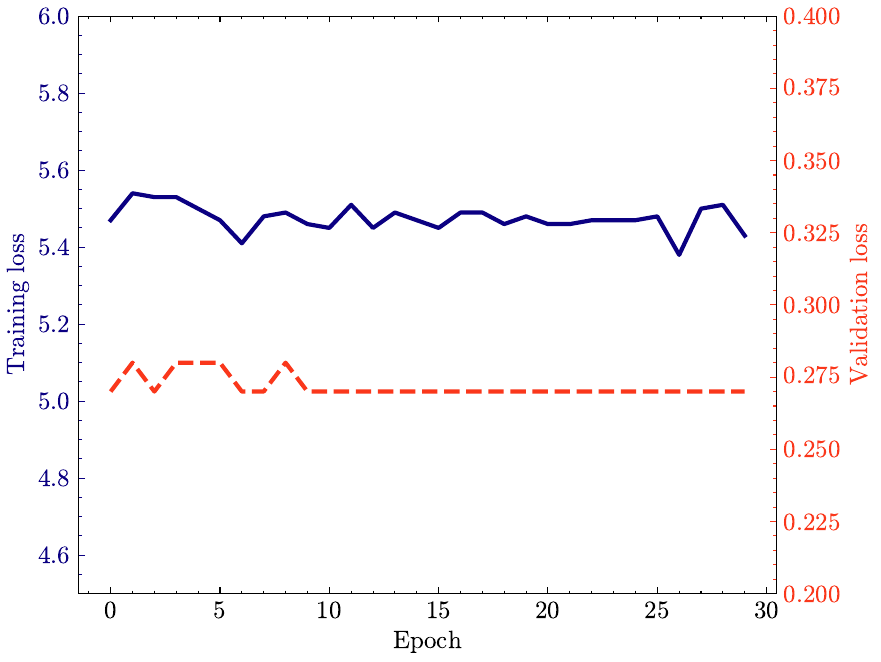}
	\end{minipage}
 \caption{The learning curves for model pre-training and fine-tuning. The left figure represents the pre-training of the ResNet-MSCFM model on the source dataset. The right figure represents fine-tuning of the pre-trained model on the target dataset. The early stopping mechanism leads to variations in the number of training epochs for the model.}
 \label{fig:learning_curve}
\end{figure}

Both the GECAM and Fermi satellites possess wide fields of view for observing GRBs. 
In typical observation scenarios without obstruction, the same burst may be detected by both satellites. 
However, there were 122 GRBs in the years 2022 and 2023 that were present in the 
Fermi burst catalog but absent from the GECAM catalog. 
To analysis this discrepancy, we extracted the light curves observed 
by GECAM at the corresponding trigger times for these bursts.
Subsequently, we applied the best-performing model trained on the GECAM dataset 
to identify the missing burst events. 
To validate the accuracy of the identified bursts, a manual assessment was conducted, 
incorporating factors such as satellite positions, earth occultation, and energy spectra. 
This comprehensive assessment process aims to determine the authenticity of the 
identified bursts and ensure the reliability of the classification results.

\begin{figure}[h]
	\begin{minipage}{\linewidth}
		\vspace{0pt}
		\centerline{\includegraphics[width=\textwidth]{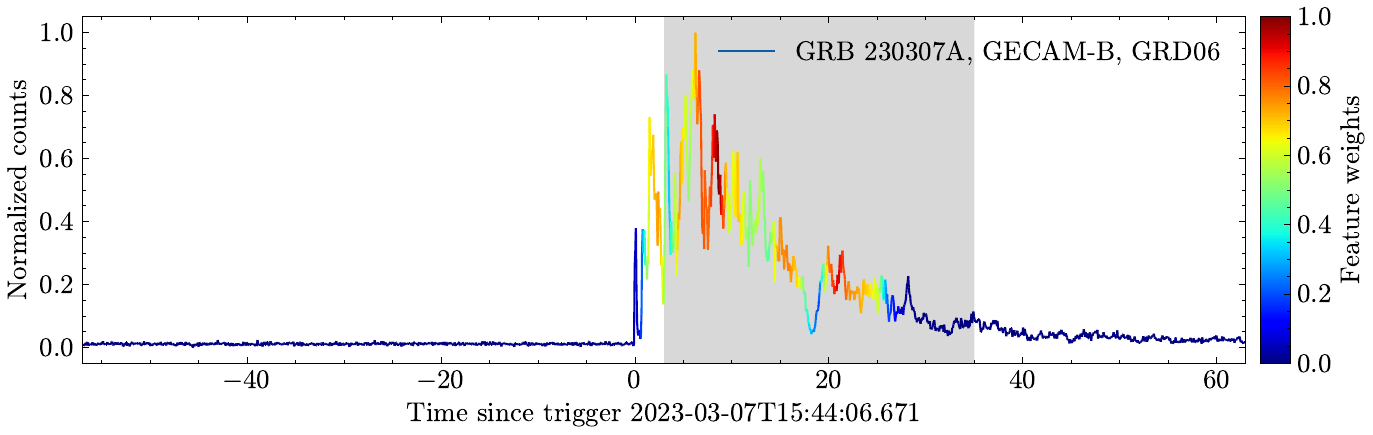}}
		\centerline{\includegraphics[width=\textwidth]{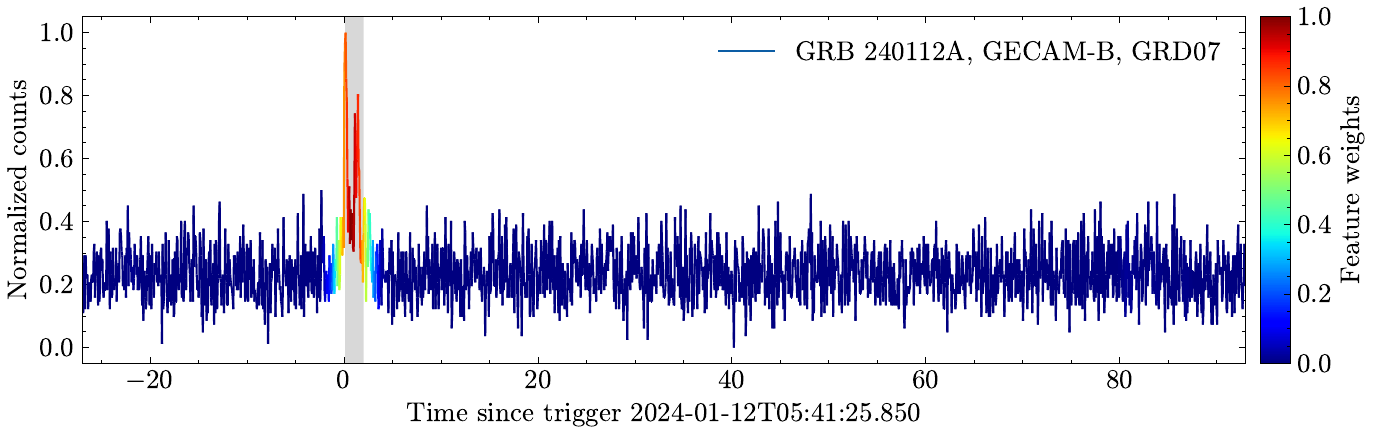}}
	\end{minipage}
	\caption{Feature visualization of two representative GRBs. The long GRB with complex structure (GRB 230307A, GRD06, \cite{230307A}) and short GRB (GRB 240112A, GRD07, \cite{240112A}) are represented. The gray area indicates $T_{90}$.}
	\label{fig:feature_cam}
\end{figure} 

\section{Result}
\label{sec:result}
The target and source datasets utilized in our study were constructed using GRBs 
detected by the GECAM-B and Fermi/GBM, respectively, as shown in Table \ref{table:dataset}. 
Light curves with a time bin of 64\,ms were extracted and binned into 9 energy bands 
to capture the photon distribution characteristics specific to GRBs.
To enhance the diversity and richness of burst signals within the datasets, 
we proposed a data augmentation method, the efficacy of which is visually demonstrated in Figure \ref{fig:crop_example}. 
This augmentation technique significantly increased the number of training samples 
for the GRB categories in the target and source datasets, expanding the samples 
from 435 to 10,005 and from 6,189 to 105,213, respectively.
Following a comparative analysis of data pre-processing methods detailed in Table \ref{table:data_preprocess}, 
we opted to standardize the light curve for each energy band. 
This standardized approach ensures uniform representation of the data across 
different energy bands, facilitating consistent model training and evaluation.

We employed a 1D convolutional neural network to classify samples belonging to 
GRB and non-GRB categories in a supervised learning framework.
To enhance feature extraction capabilities, we integrated a multi-scale feature 
cross fusion module (MSCFM) into the commonly used ResNet model. 
The hyper-parameter choices for the model are listed in Table \ref{table:hyper_parameter_select}. 
Table \ref{table:model_perform_gbm} presents the performance comparison between 
the baseline model and the model incorporating the MSCFM module on the source dataset. 
The results demonstrate that the inclusion of the MSCFM module significantly enhances the model's performance. 
Specifically, the ResNet+MSCFM model, pre-trained on the source dataset, 
achieved an impressive accuracy of 97.37\%.
Furthermore, Table \ref{table:model_perform_gecam} showcases the performance metrics 
of the model trained directly on the target dataset, solely on the source dataset, 
and through transfer learning to the target dataset. 
Notably, by transferring the pre-trained model, the model achieved a high accuracy 
of 96.41\% on the target test set.
The learning curves for model pre-training and fine-tuning are illustrated in Figure \ref{fig:learning_curve}, 
providing insights into the model's training progress.
Additionally, through feature visualization techniques, we observed that the model effectively 
extracted and focused on the key features of GRBs, as depicted in Figure \ref{fig:feature_cam}. 
These findings illustrate the model's ability to discern and focus on essential features 
for accurate GRB identification.

\begin{figure}
\centering
\includegraphics[width=\textwidth, angle=0]{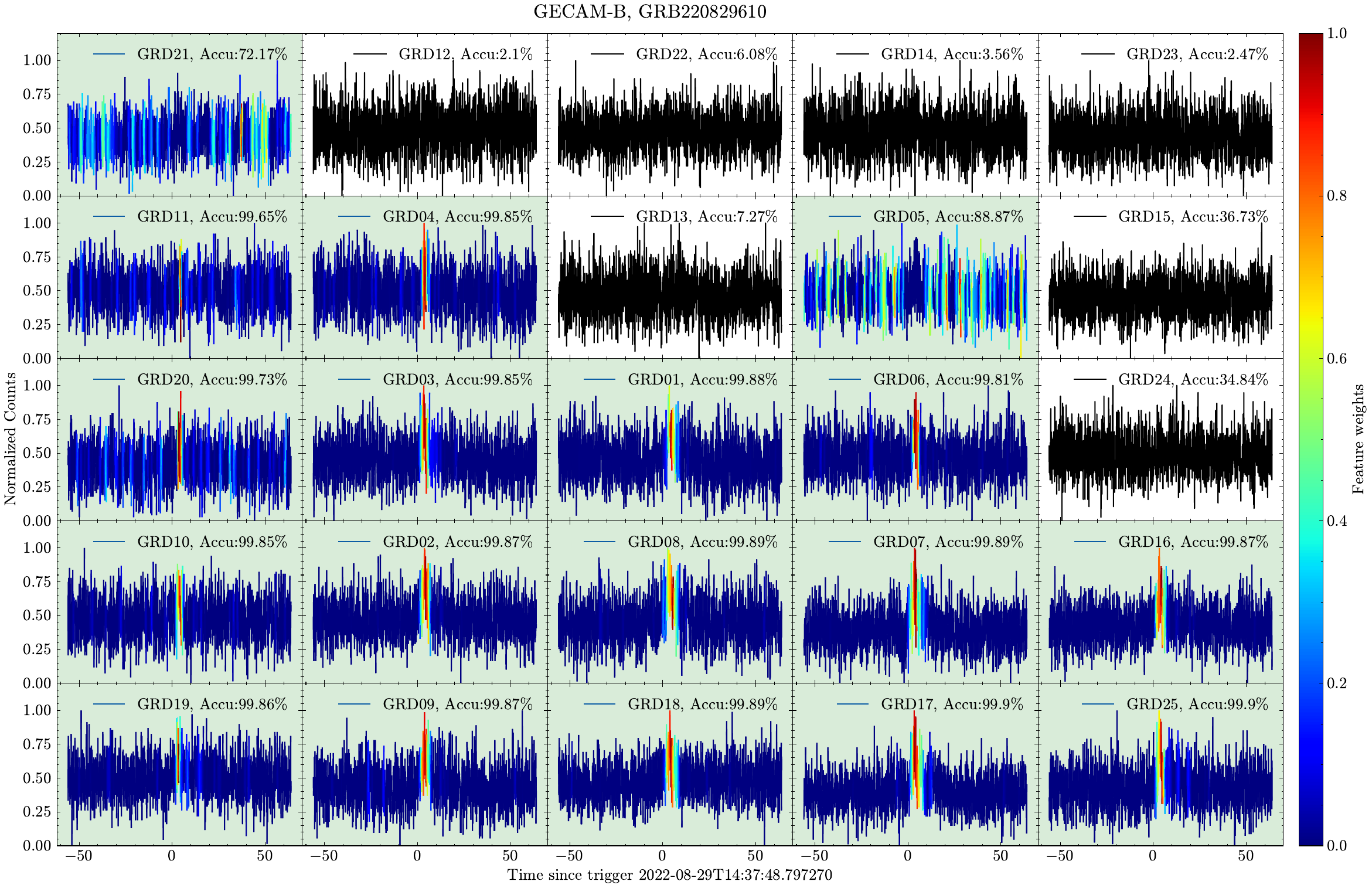}
\caption{Recovered GECAM-B GRB matched with the GRB220829610 of Fermi/GBM. The arrangement of the sub-graphs is based on the direction of the detectors. The green background indicates that the model predicts an burst signal during this time. The color represents the specific burst characteristics that the model focuses on.}
\label{fig:recover_gecam_grb_GRB220829610}
\end{figure}

\begin{figure}
\centering
\includegraphics[width=\textwidth, angle=0]{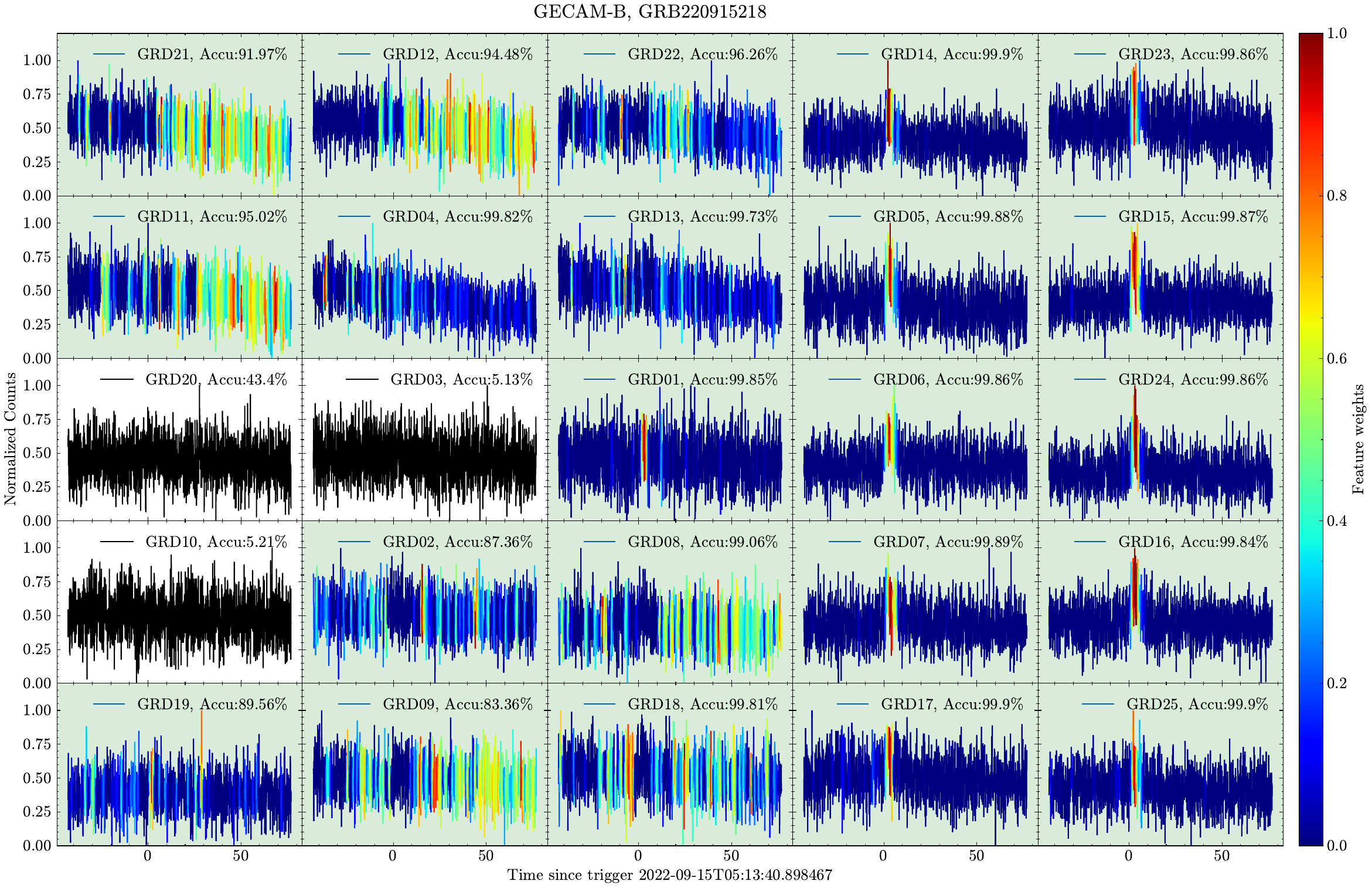}
\caption{Recovered GECAM-B GRB matched with the GRB220915218 of Fermi/GBM.}
\label{fig:recover_gecam_grb_GRB220915218}
\end{figure}

\begin{figure}
\centering
\includegraphics[width=\textwidth, angle=0]{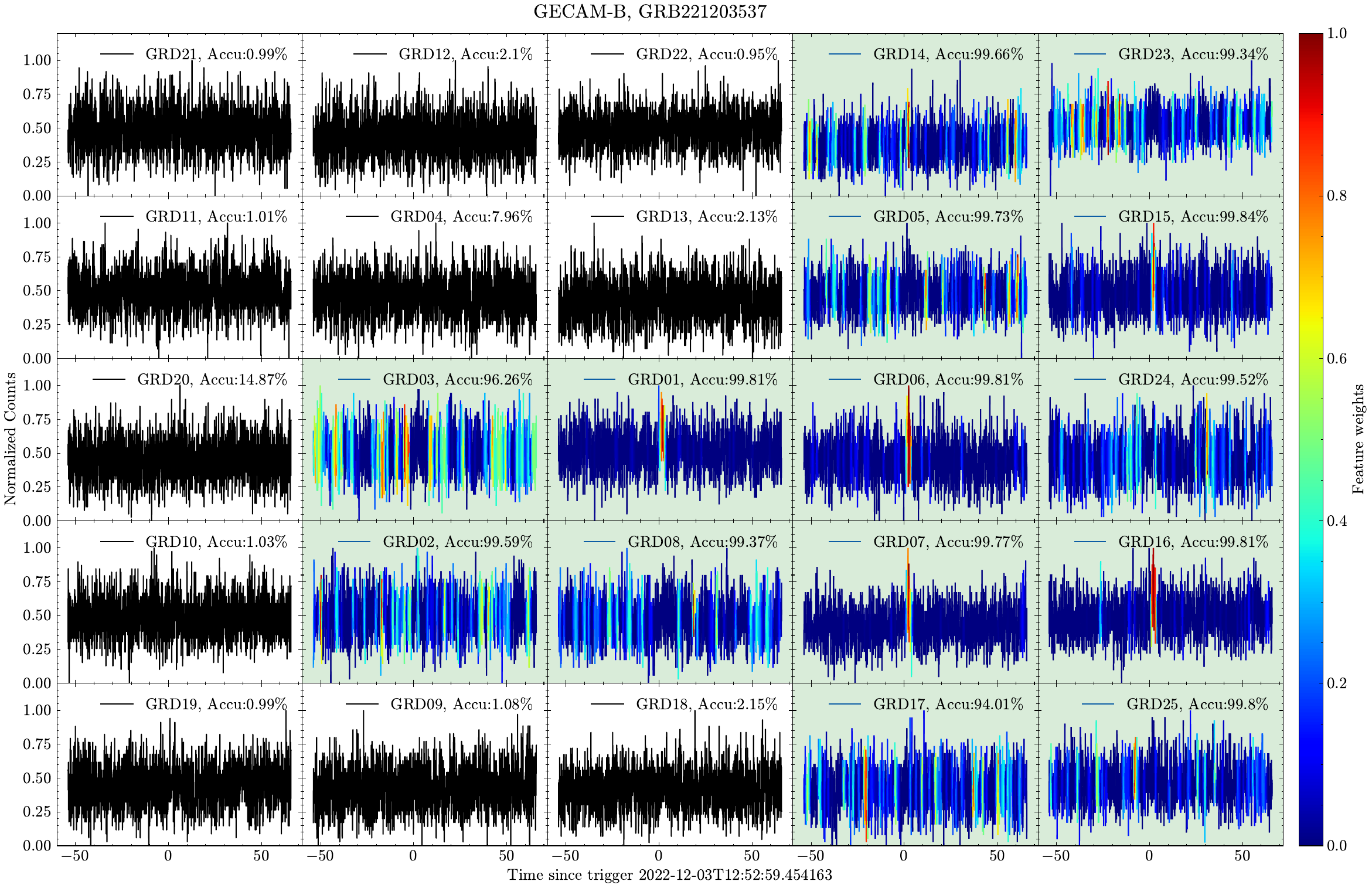}
\caption{Recovered GECAM-B GRB matched with the GRB221203537 of Fermi/GBM.}
\label{fig:recover_gecam_grb_GRB221203537}
\end{figure}

We utilized the model with the highest performance on the target dataset to identify 
previously undiscovered GRBs in the GECAM observations. 
To establish a baseline for comparison, we selected GRBs detected by Fermi/GBM in 2022 and 2023, 
excluding those already identified by the GECAM trigger search algorithm. 
The light curves of GECAM corresponding to these burst periods were extracted 
using a sliding window approach. 
Through a combination of manual analysis, we successfully recovered three GRBs, 
as depicted in Figure \ref{fig:recover_gecam_grb_GRB220829610}, 
Figure \ref{fig:recover_gecam_grb_GRB220915218}, and Figure \ref{fig:recover_gecam_grb_GRB221203537}. 
Visualizations of the extracted features revealed that the most prominent burst 
signals detected by more than two detectors align closely with the GRBs identified by Fermi/GBM. 
This consistency in signal characteristics across multiple detectors further validates 
the effectiveness of our model in accurately identifying and recovering GRBs in the GECAM observations.

\section{Discussion and Conclusion}
\label{sec:discussion} 
In traditional analytical methods, researchers strive to swiftly and precisely extract 
crucial information from GRBs by analyzing their light curves and spectra. 
In this work, we utilized supervised deep learning techniques to detect GRBs from GECAM-B observations. 
Our primary objective was to investigate the synergistic effects of data augmentation 
methods and transfer learning in enhancing the efficiency of GRB recognition, 
particularly when dealing with constraints imposed by limited observation samples.

In Table \ref{table:dataset}, we have built target and source datasets derived 
from GECAM and Fermi/GBM observations, respectively. 
To address the challenge posed by a limited sample size of GRBs, 
we opted to bin the energy bands of the light curves to align with the 
energy bands commonly associated with generic burst identifications. 
This alignment facilitates the seamless transfer of models trained on 
large-scale data, enhancing the model's adaptability to new datasets. 
Moreover, our proposed data augmentation method has proven to be instrumental 
in augmenting the diversity of burst samples significantly. 
By reducing the prominence of the burst signal while maintaining a consistent 
background level, as exemplified in Figure \ref{fig:crop_example}, 
this method has effectively expanded the number of training samples 
from a modest few thousand to a substantial hundred thousand. 
The integration of data augmentation techniques has mitigated the challenge 
of limited training samples in deep learning applications for GRB identification, 
thereby improving the model's generalization capabilities.

We selected ResNet as the benchmark model for GRB identification due to its robustness 
and versatility in deep learning applications. 
To enhance the model's multi-scale feature extraction capabilities, 
we introduced a multi-scale feature cross fusion module (MSCFM). 
The detailed structure of the model, including the implementation of the MSCFM module, 
is illustrated in Figure \ref{fig:network_architectures}. 
Table \ref{table:hyper_parameter_select} provides a comprehensive evaluation 
of hyper-parameter selection, assessing the impact of hyper-parameters and optimizing the model.
Our approach involved pre-training the model on the source dataset, as shown in Table \ref{table:model_perform_gbm}. 
Incorporating the MSCFM module led to a noticeable improvement in the accuracy of the ResNet model, 
indicating that the fusion of multi-scale features enabled the model to capture 
richer information and enhance its generalization capacity.
Table \ref{table:model_perform_gecam} presents the performance comparison of models with 
and without pre-training and fine-tuning. 
We observed that pre-trained models, particularly those trained on extensive datasets, 
exhibited superior generalization abilities. 
Fine-tuning the parameters of pre-trained models on the target dataset 
further enhanced their adaptability, highlighting the efficiency of 
transfer learning in handling limited data scenarios.
The learning curve depicted in Figure \ref{fig:learning_curve} demonstrates 
the model's effective training process, 
with the early stopping mechanism preventing over-fitting. 
Additionally, the visualization of key features in Figure \ref{fig:feature_cam} 
showcases the model's ability to focus on essential burst characteristics, 
aligning with human expertise in GRB recognition.
Furthermore, Figures \ref{fig:recover_gecam_grb_GRB220829610}, \ref{fig:recover_gecam_grb_GRB220915218}, and \ref{fig:recover_gecam_grb_GRB221203537} 
illustrate the successful recovery of three GRBs in real GECAM-B 
observations using the optimized model. T
his achievement can be attributed to the synergistic effects of our data augmentation 
algorithms, model enhancements, and transfer learning strategies.
The accurate identification of GRBs holds significant importance, as they may potentially 
correspond to gravitational waves, fast radio bursts, or other transient 
astronomical phenomena \citep{sub_grb_for_GW}.

Transfer learning shows as a valuable strategy for addressing the challenge of 
limited data samples, thereby enhancing the generalization capacity of deep learning models. 
The integration of a 1D convolutional neural network with a multi-scale feature cross 
fusion module yielded notable performance improvements following pre-training on a comprehensive dataset. 
This enhancement not only bolstered the model's generalization capabilities but 
also facilitated seamless adaptation to GECAM data. 
This is demonstrated by the model's precise identification of burst events, 
in alignment with expert human analysis.
In the future, we plan to incorporate this refined methodology into the GECAM data 
analysis pipeline to streamline the identification of GRBs. 
Furthermore, the scalability of this approach allows for potential extension to other 
satellite missions in the future, such as GECAM-C (HEBS, \cite{gecam_c}),
GECEM-D (DRO/GTM, \cite{GECAM_d}), HXMT \citep{hxmt}, and SVOM/GRM \citep{svom}. 
By leveraging data from multiple satellites, a collaborative multi-satellite analysis 
could be pursued to enhance GRB searches, potentially enabling follow-up observations 
across various wavelength bands. 
This collaborative effort holds promise for capturing the elusive early afterglow 
of GRBs, underscoring the significance of interdisciplinary cooperation in advancing 
our understanding of these transient astronomical phenomena.

\normalem
\begin{acknowledgements}
We would like to thank Prof XXX,Prof. Rui Luo, Dr. Crupi, Riccardo and 
Dr. Yi Yang for helpful discussion. 
We thank Sheng-lun Xie, Yue Wang for sharing  of data analyzing.
This study is supported by the National Natural Science Foundation 
of China (grant Nos. 12103055, 12273042, 12133007, 41827807, and 61271351) and 
by the National Key R\&D Program of China (2021YFA0718500). 
This work is also partially supported by the Strategic Priority Research 
Program of the CAS under grant No. XDA15360300. B. L. acknowledges support 
from the National Astronomical Science Data Center Young Data Scientist 
Program (grant No. NADC2023YDS-04). 
\end{acknowledgements}

\bibliographystyle{raa}
\bibliography{paper}

\end{document}